\documentclass[aps,twocolumn,showpacs]{revtex4}
\usepackage{graphicx}
\usepackage{bm}
\usepackage{amsmath}

\begin{document}

\title{Helical edge and surface states in HgTe quantum wells and bulk insulators}
\author{Xi Dai$^{1,3}$, Taylor L. Hughes$^{2}$, Xiao-Liang Qi$^{2}$, Zhong Fang$^{1}$ and Shou-Cheng Zhang$^{2}$}

\affiliation{$^1$ Beijing National Laboratory for Condensed Matter
Physics, Institute of Physics, Chinese Academy of Sciences,
Beijing, 100080, China}

\affiliation{$^2$ Department of Physics, McCullough Building,
Stanford University, Stanford, CA 94305-4045}

\affiliation{$^3$ Department of Physics, University of Hong Kong,
Pokfulam Road, Hong Kong}

\begin{abstract}
The quantum spin Hall (QSH) effect is the property of a new state
of matter which preserves time-reversal, has an energy gap in the
bulk, but has topologically robust gapless states at the edge.
Recently, it has been shown that  HgTe quantum wells realize this
novel effect\cite{bernevig2006d}. In this work, we start from
realistic tight-binding models and demonstrate the existence of
the helical edge states in HgTe quantum wells and calculate their
physical properties. We also show that  $3$d HgTe is a topological
insulator under uniaxial strain, and show that the surface states
are described by single-component massless relativistic Dirac
fermions in $2+1$ dimensions. Experimental predictions are made
based on the quantitative results obtained from realistic
calculations.

\end{abstract}


\maketitle

Conventional insulators have a gap for all charge excitations and
their physical properties are not sensitive to changes in the
boundary conditions. Recently, a new class of quantum spin
Hall(QSH) insulators has been proposed theoretically in both 2d
and
3d\cite{kane2005A,bernevig2006A,bernevig2006d,fu2006A,moore2006}.
The QSH insulators are invariant under time reversal, have a
charge excitation gap in the bulk, but have topologically
protected gapless edge states that lie in the bulk insulating gap.
This type of insulator is typically realized in spin-orbit coupled
systems; and the corresponding edge states have a distinct helical
property: states with one spin-polarization move around the sample
edge in one direction, while states with the opposite
spin-polarization move in the opposite
direction\cite{kane2005A,wu2006,xu2006}. The helical edge states
are responsible for the intrinsic spin Hall effect in the
insulating
state\cite{murakami2003,sinova2004,murakami2004,murakami2004a}.
The edge states come in Kramers's doublets, and time reversal
symmetry ensures the crossing of their energy levels at special
points in the Brillouin zone. Because of this level crossing, the
spectrum of a QSH insulator cannot be adiabatically deformed into
that of a topologically trivial insulator without helical edge
states; therefore, in this precise sense, the QSH insulators
represent a topologically distinct new state of matter. In $3$d
the class of strong topological insulators has a similar
distinction\cite{fu2006A,fu2006B} and is the natural
generalization of the QSH insulator.

The study of the QSH effect in quasi-$2$d ${\rm HgTe}/{\rm CdTe}$
quantum wells carried out in Ref.\cite{bernevig2006d} is based on
a simplified model obtained by ${\bf k\cdot P}$ perturbation
theory and the envelope function approximation. The conclusion of
a topological quantum phase transition is reached based on a
Dirac-type subband level crossing at the $\Gamma$ point. However,
such a level crossing is not generic and could be avoided in the
real system due to the bulk inversion asymmetry (BIA) of the
zinc-blende lattice. Consequently, a more realistic study is
necessary to obtain a better understanding of the QSH phase and
the topological phase transition. In this paper, we study the
subband structure and edge state properties of ${\rm HgTe}/{\rm
CdTe}$ quantum wells using a realistic tight-binding(TB) model.
The level crossing avoided at the $\Gamma$ point is recovered at
several finite wavevectors. In other words, the phase transition
between two different insulating regions remains robust despite
inversion symmetry breaking. Furthermore, the topological nature
of the QSH regime is demonstrated explicitly by studying the
properties of the helical edge states in an open boundary system.
We also apply the same realistic TB calculations to uniaxial
strained $3$d ${\rm HgTe}$, and obtain the topologically
non-trivial surface states. Thus, strained bulk ${\rm HgTe}$ is
demonstrated to be a strong topological insulator. Ref.
\cite{fu2006B} investigates a class of models with bulk inversion
symmetry which strictly speaking does not apply to ${\rm HgTe}$,
however, adiabatic continuity is used to argue that strained bulk
${\rm HgTe}$ could be a strong topological insulator.

%\section{Tight-binding model and Green function method}

The CdTe and HgTe materials have the same zinc-blende lattice
structure and are well-described by the same type of tight binding
Hamiltonian, albeit with different parameters. This Hamiltonian
includes two s-type orbitals and three p-type orbitals on each
atom and reads
\begin{widetext}
\begin{eqnarray}
H=\sum_{i\sigma \overrightarrow{R}}E_{i,a}a_{i\sigma \overrightarrow{R}%
}^{\dag }a_{i\sigma \overrightarrow{R}}+\sum_{i\sigma \overrightarrow{R}%
}E_{i,c}c_{i\sigma ,\overrightarrow{R}+\overrightarrow{d}}^{\dag
}c_{i\sigma ,\overrightarrow{R}+\overrightarrow{d}}
+\sum_{\overrightarrow{R},\overrightarrow{R^{\prime }},\sigma
ij}V_{i,j}a_{i\sigma \overrightarrow{R}}^{\dag }c_{i\sigma ,\overrightarrow{R%
}+\overrightarrow{d}}+H.C. \nonumber\\
+\sum_{ij\sigma \sigma ^{\prime }\overrightarrow{R}%
}\frac{4\lambda _{a}}{\hbar }\overrightarrow{L}_{a,ij}\cdot \overrightarrow{S%
}_{a,\sigma \sigma ^{\prime }}a_{i\sigma \overrightarrow{R}}^{\dag
}a_{j\sigma ^{\prime }\overrightarrow{R}}
+\sum_{ij\sigma \sigma ^{\prime }\overrightarrow{R}}\frac{4\lambda _{c}}{%
\hbar }\overrightarrow{L}_{c,ij}\cdot \overrightarrow{S}_{c,\sigma
\sigma ^{\prime }}c_{i\sigma
\overrightarrow{R}+\overrightarrow{d}}^{\dag }c_{j\sigma ^{\prime
}\overrightarrow{R}+\overrightarrow{d}}\label{TBHamiltonian}
\end{eqnarray}
\end{widetext}
\noindent where $a_{i\sigma \overrightarrow{R}}^{\dag }$ and $c_{i\sigma ,%
\overrightarrow{R}+\overrightarrow{d}}^{\dag }$ are the creation operators
for electrons on the \emph{anion} and \emph{cation} sites respectively,  $%
i=(s,s^{\ast },p_{x},p_{y},p_{z})$\  is the orbital index,  and $\sigma $ is the spin index. $%
E_{i,a},E_{i,b}$ and $V_{i,j}$ are the tight binding parameters
defined by Slater and Koster\cite{Slater1954}. Spin-orbit coupling
is contained in the last two terms, and is represented by two
coupling constants $\lambda _{a}$ and $\lambda _{b}$. The tight
binding parameters are taken from Ref. \cite{Kobayashi1982} where
they were determined by fitting to a first-principles calculation.

To obtain the boundary (edge or surface) states for quasi-$2$d
HgTe/CdTe quantum well structures and for the bulk $3$d materials,
we apply the Green's function method\cite{Turekbook,Bryant1987}
based on the TB model described above. For the quantum well
system, we consider a symmetric HgTe/CdTe hetero-structure with a
fixed $N_c=8$ layers of CdTe surrounding a variable $N_h$ layers
of HgTe on each side. In order to calculate the boundary states we
choose open boundary conditions along the $x$-direction, periodic
boundary conditions along the $y$-direction, and the open boundary
condition along the $z$-direction which is the growth direction of
the quantum well. For the bulk materials, surface states can be
calculated for a semi-infinite system with open boundary
conditions in one direction, such as $[001]$, and periodic
boundary conditions in the other two directions.
%{\em (To Xi: Here and below, could you please replace
%the x, y, z label to the standard labels like $[001]$, etc.? or add
%an explanation to the choice of x y z axis.)}
The inverse Green's functions of both the quantum well and bulk
materials can be written in a block tri-diagonal form as
\begin{equation*}
G^{-1}\left( z\right) =z-H=\left(
\begin{array}{cccc}
z-H_{0} & C & 0 & 0 \\
C^{\dag } & z-H_{0} & C & 0 \\
0 & C^{\dag } & z-H_{0} & C \\
0 & 0 & C^{\dag } & \cdots%
\end{array}%
\right)
\end{equation*}
\noindent where the diagonal block $H_{0}$ describes the
Hamiltonian within the same ``principal layer",\cite{Turekbook}
that is, the layer along the x-direction with open boundary
conditions for the quantum well case, or that along the only
direction with open boundary condition for the bulk case. The
off-diagonal block $C$ describes the coupling between two
nearest-neighbor principal layers. To study the boundary states we
only need $g_{ij}$, the Green's function on the boundary, where
$i,j$ are the indices of the local basis on the boundary. This
function is contained in the first diagonal block of the matrix
$G\left( z\right) =\left( z-H\right) ^{-1}$ and can be expressed
in a recursive way as $g_{ij}^{(N)}=\left(
z-H_{0}-Cg^{(N-1)}C^{\dag }\right)_{ij}^{-1}$ with $g_{ij}^{(N)}$
\ denoting the boundary Green's function for a system with $N$
principal
layers. The above recursive equations can be closed by the initial condition $%
g_{ij}^{(1)}=\left( z-H_{0}\right) ^{-1},$ and we obtain
$g_{ij}^{(N)}$ iteratively. Any physical observables projected
onto the boundary are easily expressed using these Green's functions as $-\frac{1}{%
\pi }\int d\omega \sum_{ij}{Im}g_{ij}^{(N)}(\omega +i0^{+})O_{ji}=-\frac{1}{%
\pi }\int d\omega \rho _{o}\left( \omega \right) $ where $\rho
_{o}\left( \omega \right) $ is a type density of states and $i,j$
run over all the local basis states on the boundary. For example,
 $O^{c}=\sum_{i}|i><i|,$ and $O^{s}=\sum_{ij}L^z_{ij}+S^z_{ij}$ generate
 the density of states for charge and spin on the boundary
 respectively. Notice here we generalize the definition of ``spin" to
 include all the local angular momentum with real spin and angular
 momentum of the local basis.
This method is easily generalized to study the interface states
between two semi-infinite crystals, {\em e.g.}, ${\rm HgTe}$ and
${\rm CdTe}$, which is essential in the present work as will be
explained below.

%\section{${\rm HgTe}/{\rm CdTe}$ quantum well}

\begin{figure}[tbp]
\begin{center}
\includegraphics[width=8cm,angle=0,clip=]{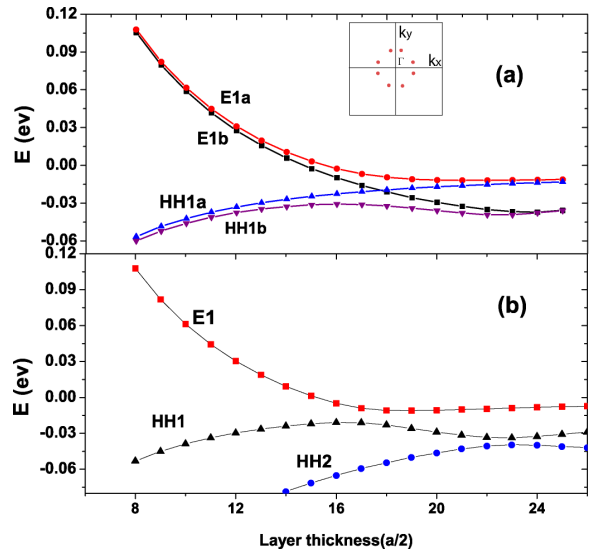}
\end{center}
\caption{(a) The subband splitting as the function of layer
thickness at one of the crossing points
$\textbf{k}=(0.017,0.008){\frac{\pi}{a}}$.(Inset) Schematic
diagram of all 8 crossing points in the 2d BZ (b)The subband
splitting as the function of layer thickness at the $\Gamma$
point.  } \label{fig1}
\end{figure}

 As is well-known for ${\rm HgTe}$ quantum wells, the confinement effect
along the $z$-direction opens a small gap around the Fermi level
which makes the quantum well an insulator. By the analysis in Ref.
\cite{bernevig2006d} based on the $\textbf{k}\cdot\textbf{P}$
approximation, a quantum phase transition from a topologically
trivial phase to a non-trivial phase (QSH phase) occurs for ${\rm
CdTe}/{\rm HgTe}$ quantum-wells at some critical thickness of
${\rm HgTe}$ layer, which is signaled by a level crossing between
the E1 and HH1 subbands. We study this transition using the more
realistic full TB Hamiltonian given above. First we choose
periodic boundary conditions in $x$ and $y$ to obtain the $2$d
subband spectrum for different thicknesses $N_h$ of ${\rm HgTe}$
layers. In Fig.\ref{fig1} (b), we plot the subband spectrum at the
$\Gamma$-point as a function of HgTe ``half"-layers. Whereas Ref.
\cite{bernevig2006d} predicts that the subbands will cross, we
find an anti-crossing at the $\Gamma$-point at a critical layer
thickness $d_{c}=9a.$ This difference between the work of
Ref.\cite{bernevig2006d} and these TB results can be explained by
the presence of BIA in the zinc-blende lattice which is ignored in
the previous paper. From the $\textbf{k}\cdot\textbf{P}$
perspective, an additional term $H'=C_k
k_z\{J_z,(J_{x}^2-J_{y}^2)\}$ is allowed in the bulk Hamiltonian
once the point group symmetry is reduced by BIA to
$D_{2d}$\cite{winklerbook}, with $J_x, J_y, J_z$ the spin-$3/2$
matrices. Despite being a $k_z$-dependent term in the bulk system,
in the quantum well $H'$ generates a constant term with
\emph{finite} matrix elements connecting the $|E1\pm>$ subbands
with the $|HH1\mp>$ subbands near the $\Gamma$-point, which can be
derived following the effective $2$d $\textbf{k}\cdot\textbf{P}$
approach of Ref.\cite{bernevig2006d}. This term qualitatively
affects the physics in exactly the manner predicted by the TB
model.

Upon further investigation, we find that the inversion asymmetry
shifts the crossing point from the $\Gamma$-point to eight
non-zero $k$-points around it. In Fig.\ref{fig1} (a), we plot the
subband spectrum at one of these crossing points
$\textbf{k}=(0.017,0.008){\frac{\pi}{a}}$ where a clear level
crossing between $E1$ and $HH1$ is observed. (Notice that away
from $\Gamma$-point, the inversion asymmetry removes the double
degeneracy as expected.) The other seven crossing points are
determined by the point-group star of this one. We find that even
though the level crossing at $\Gamma$ point is avoided, the
gap-closing quantum phase transition between two insulating
regimes predicted previously in the simplified
model\cite{bernevig2006d} still exists in the more realistic case.

\begin{figure}[tbp]
\begin{center}
\includegraphics[width=9cm,angle=0,clip=]{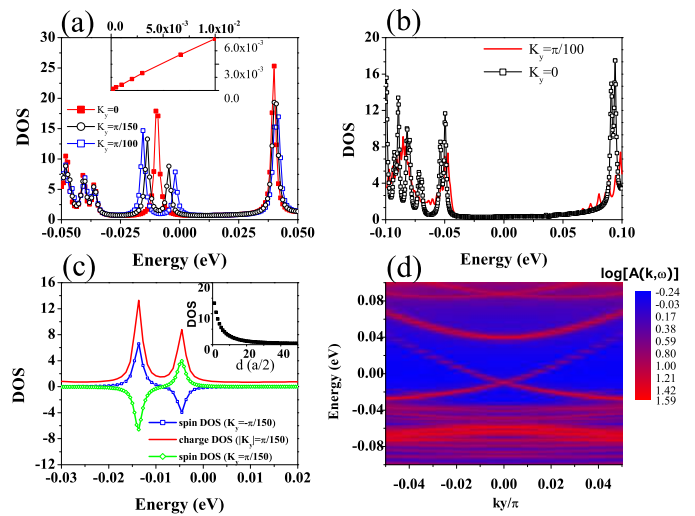}
\end{center}
\caption{ (a)The density of states at the edge of the quantum well
with layer thickness $d=20(a/2)$. (The inset plot shows the linear
energy splitting of the edge states in the very small region near
$k_y=0$.)(b)The density of states at the edge of the quantum well
with layer thickness $d=10(a/2)$. (c) The spin density of states at
the edge of the quantum well with thickness $d=20(a/2)$. (The inset
plot shows the decay of density of states at the peak energy of the
spectra with $k_y=0$.)(d)The intensity color plot on the
energy-momentum plane for the density of states at the edge of the
quantum well with thickness $d=20(a/2)$. } \label{fig2}
\end{figure}

After identifying the presence of the quantum phase transition,
the next natural question is whether the topologically non-trivial
phase on the thicker side of the transition predicted by the
$\textbf{k}\cdot\textbf{P }$ calculation will survive the
inversion symmetry breaking. The most convincing way to answer
this question is to calculate the edge states directly. By the
recursion method we obtain the charge and spin density of states
defined above. The charge density of states for a quantum well
structure with $20$ and $10$ ``half"-layers are plotted in
Figs.\ref{fig2} (a) and (b) respectively. Sharp peaks appear in
the gap for the quantum well with 20 half-layers and are absent in
the $10$ half-layer system. As required by time reversal symmetry
(TRS), the energy levels at the $\Gamma$ point \emph{must} be
doubly degenerate, but can be split at finite $k_y$. The energy
level splitting near $k_y=0$ as a function of $k_y$ is plotted in
the inset of Fig. \ref{fig2}(a), which shows a perfect linear
dispersion indicating a level crossing of the edge states at
$k_y=0.$  The spin densities of states for two $k_y$'s with
opposite signs are plotted in Fig. \ref{fig2}(c). If the chemical
potential lies between the two peaks, only the lower branch of the
edge states is occupied and a spin current will be carried by the
edge states. With the recursion method we can also obtain the
charge density of states on the inner layers away from the edge.
In the inset of Fig. \ref{fig2}(c), we plot the height of the
in-gap peak for $k_y=0$ on the different layers, which decays very
fast from the edge and thus demonstrates that the in-gap peak is
produced by the edge states. Finally, we plot the edge state
dispersion in Fig. \ref{fig2}(d) with a color intensity plot
generated from the density of states. We find that the edge states
merge with the bulk states very quickly as $k_y$ moves away from
the $\Gamma$-point. Consequently, whenever the chemical potential
lies in the bulk gap, the only low-energy states crossing the
Fermi level are {\em one} Kramers's pair of edge states. In other
words, the low energy behavior of the quantum well in the bulk
insulating region is described by an odd number of pairs of $1$d
channels propagating on each edge. According to
Ref.\onlinecite{wu2006}, such a $1$d liquid is a ``helical
liquid," and cannot be realized in any pure $1$d system that
preserves TRS. It can only exist as the edge theory of a $2$d QSH
insulator. In this way, the results of our calculations provide
convincing evidence that the HgTe/CdTe quantum well is a QSH
insulator characterized by a non-trivial $Z_2$ topological
invariant\cite{bernevig2006d,kane2005A}.

%\section{bulk ${\rm HgTe}$ and ${\rm HgTe}/{\rm CdTe}$ interface}

The same recursion method can also be applied for the $2$d surface
states of strained $3$d ${\rm HgTe}$, which was recently suggested
to be topologically nontrivial\cite{fu2006B}. However, the TB
model (\ref{TBHamiltonian}) applied to a semi-infinite system with
a $2$d surface always generates some surface states, even for a
trivial band insulator, {\em e.g.}, ${\rm CdTe}$.\cite{Bryant1987}
Physically, these surface states correspond to a dangling
$sp^3$-hybridized bond at each surface atom, and are thus strongly
dependent on the details of the surface physics, such as surface
reconstruction and disorder. Since in the present paper we are
only concerned about the topological properties, which are
insensitive to the details of surface physics, we can choose a
surface regularization that removes the trivial surface states and
leaves only the topological ones. Therefore, we will focus on the
interface between bulk ${\rm HgTe}$ and ${\rm CdTe}$, where the
dangling bonds of ${\rm HgTe}$ are coupled to ${\rm CdTe}$ so the
trivial surface states vanish. Since ${\rm CdTe}$ can be
adiabatically connected with vacuum by taking its band gap to
infinity, the topological properties of ${\rm HgTe}/{\rm CdTe}$
interface are determined by ${\rm HgTe}$.

Since bulk ${\rm HgTe}$ is a semi-metal, we need to apply a small
compressive strain, along say the $[001]$ direction, to make it an
insulator\cite{murakami2004a}. The tight binding parameters for
the strained HgTe are obtained by fitting the LDA results from the
plane wave pseudo potential method\cite{Fang2002}. In the present
paper, we apply the constrain along the $[001]$ direction
(z-direction) and locate the $HgTe/CdTe$ interface along both the
$[100]$ (x-direction) and $[001]$ directions. And for both cases,
we find non-trivial surface states.
%Our Green's function calculation shows that both the HgTe and CdTe
%have strong surface states in the gap. Although the existence of
%the surface state in CdTe and HgTe has been known for a few
%decades\cite{hgtecdtesurfacstates}, their topological structure
%has never been studied carefully.
%Although the
%surface states in HgTe and CdTe look quite similar there is an
%essential topological difference between them. The surface states
%for CdTe have level crossings at all the high symmetry points:
%$(0,0),(\pi,\pi),(2\pi,0),(0,2\pi)$. In contrast, the HgTe surface
%states merge into the bulk bands near the $\Gamma$-point (as shown
%in the inset of fig.\ref{fig3}), but behave similar to CdTe in the
%rest of the Brillouin zone.
In Fig. 3 we plot the charge density of states in a color
intensity plot. The single pair of surface states is clearly seen
in the bulk insulating gap and they cross at the $\Gamma$-point.
There are no other surface states in the entire zone. Any pure
$2$d band theory that respects TRS must have the same number of
Kramers's pairs on each of the time-reversal symmetric wave
vectors because all of the $2N$ energy bands must be paired on
these wavevectors. As a result, the $2$d surface states of HgTe
cannot emerge from any pure $2$d surface effect, and only as a
consequence of bulk topology.

\begin{figure}[tbp]
\begin{center}
\includegraphics[width=2.5in,angle=0,clip=]{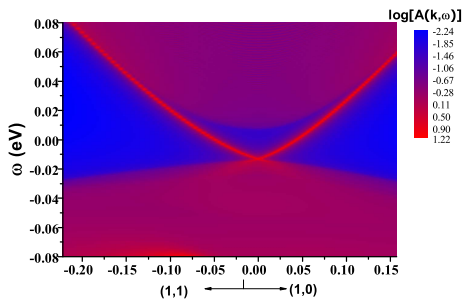}
\end{center}
\caption{Intensity color plot in the energy-momentum plane for the
density of states at the ${\rm HgTe}-{\rm CdTe}$ interface. The
uniaxial strain is applied along the $[001]$ direction by choosing
the $c/a$ ratio to be $0.98$ and the $HgTe/CdTe$ interface is
chosen along $[100]$ direction. } \label{fig3}
\end{figure}

\begin{figure}[th]
\begin{center}
\includegraphics[width=3.2in] {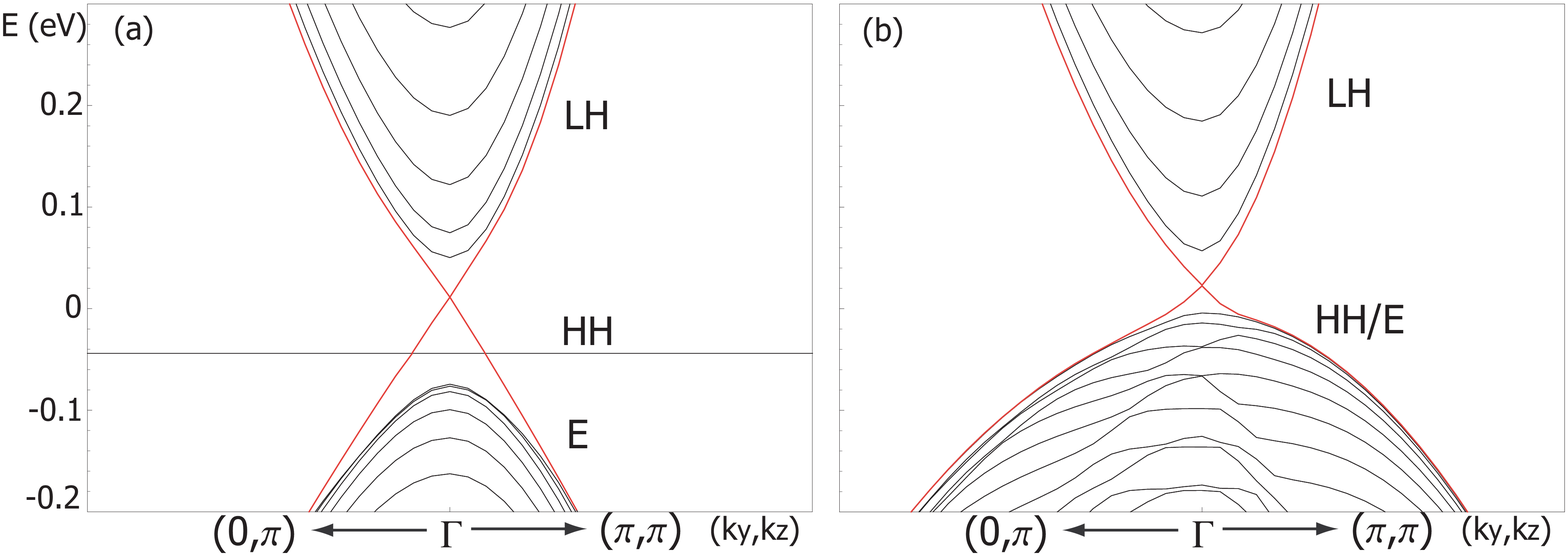}
\end{center}
\caption{Band structure near the $\Gamma$-point for (a) Decoupled
HH band (b)Full HH band coupling. Surface sates are shown in red.
Strain induced gap is artificially large so that surface states
are clearly visible. Generally, topologically non-trivial surface
states exist for any finite compressive strain.} \label{3dhgtekp}
\end{figure}

It is useful to try to understand this result from a continuum
$\textbf{k}\cdot\textbf{P}$ perspective. HgTe has a non-trivial
topological structure because the band-structure is \emph{only}
``inverted" near the $\Gamma$-point. The fact that an occupied
band at this point has $\Gamma^6$ character means that the $Z_2$
invariant picks up an extra factor of $(-1)$ (if we ignore the
small BIA) making it non-trivial. Due to strong orbital mixing the
$\Gamma^6$ character is washed out as one moves away from the
$\Gamma$-point and at the other special TR invariant points the
``inverted" structure is absent. Therefore, we should be able to
understand the topological properties from the band structure
\emph{only} near the $\Gamma$-point. The key point is to consider
the full $6$-band Kane model\cite{kane1957} instead of just the
reduced $4$-band Luttinger model\cite{sham1985}. If we only keep
the bands in the Luttinger model the topological structure is
absent. In the presence of uniaxial compressive strain along the
(001) direction an insulating gap opens between the heavy-hole(HH)
and light-hole(LH) bands by pushing the HH band downward in
energy. For a moment we will ignore the HH band and focus only on
the LH and $\Gamma^6$ (E) band. From the form of the Kane model,
the coupling of the LH and E bands near the $\Gamma$-point is
exactly a $3$d anisotropic massive Dirac Hamiltonian if we ignore
BIA and keep the leading order in $k$. The Dirac Hamiltonian
preserves parity symmetry and we can label the bands by parity
eigenvalues. Since the coupling is linear there must be one even
(doubly degenerate) and one odd (doubly-degenerate) band. We
expect that when the odd parity band lies below the even band then
there will be a non-trivial $Z_2$ invariant which indicates an odd
number of pairs of surface states that cross at TR invariant
points\cite{fu2006B}. The presence of the HH band will change the
features of the spectrum but it does not change the presence of
the surface states, or their protected crossing, as long as the
strain induced gap is open. The system will remain a $3$d
topological insulator when the HH band is coupled, and when BIA
terms are added, as long as the bulk gap does not close.  To show
evidence of our statements we solve the $6$-band Kane model on a
cylinder. First we solve the model with the HH band
\emph{completely} decoupled from the LH and E bands (Fig.
\ref{3dhgtekp}a). Here the HH band remains flat and is split from
the LH band by the strain induced gap. In the gap there are clear,
linearly dispersing surface states which traverse the gap between
the LH and E bands. Nothing occurs at the other special points in
the BZ, therefore this is a strong topological insulator. Turning
on the coupling to the HH band changes features of the band
structure but does not change the topology of the state since the
gap between the LH and HH bands never closes. It is clear from
Fig. \ref{3dhgtekp}b that even when the HH band is fully coupled
the system is still a strong topological insulator with surface
states crossing at $\Gamma.$

This work is supported by the NSF under grant numbers DMR-0342832
and the US Department of Energy, Office of Basic Energy Sciences
under contract DE-AC03-76SF00515, the MARCO Center on Functional
Engineered Nano Architectonics (FENA), and the Knowledge
Innovation Project of the Chinese Academy of Sciences.

\bibliography{hgte}
\end{document}